\documentclass[aps,twocolumn,prl,superscriptaddress,showpacs,amsmath,amssymb]{revtex4-1}

\usepackage{graphicx}
\usepackage{dcolumn}
\usepackage{bm}
\usepackage{subfigure}
\usepackage{tabularx}
\usepackage{color}

\begin{document}

\title{Collaboration networks from a large CV database:
\\dynamics, topology and bonus impact}

\author{Eduardo B. Ara\'ujo}
\affiliation{Departamento de F\'isica, Universidade Federal do Cear\'a, 60451-970, Cear\'a, Brazil}
\author{Andr\'e A. Moreira}
\affiliation{Departamento de F\'isica, Universidade Federal do Cear\'a, 60451-970, Cear\'a, Brazil}
\author{Vasco Furtado}
\affiliation{N\'ucleo de Aplica\c{c}\~{a}o em Tecnologia da Informa\c{c}\~{a}o, Universidade de
Fortaleza, 60811-905, Cear\'a, Brazil}
\author{Tarcisio H. C. Pequeno}
\affiliation{N\'ucleo de Aplica\c{c}\~{a}o em Tecnologia da Informa\c{c}\~{a}o, Universidade de
Fortaleza, 60811-905, Cear\'a, Brazil}
\author{Jos\'e S. Andrade Jr.}
\affiliation{Departamento de F\'isica, Universidade Federal do Cear\'a, 60451-970, Cear\'a, Brazil}

\date{\today}

\begin{abstract}
Understanding the dynamics of research production and collaboration may reveal better strategies for
scientific careers, academic institutions and funding agencies. Here we propose the use of a large
and multidisciplinar database of scientific curricula in Brazil, namely, the Lattes Platform, to
study patterns of scientific production and collaboration. In this database, detailed information
about publications and researchers are made available by themselves so that coauthorship is
unambiguous and individuals can be evaluated by scientific productivity, geographical location and
field of expertise. Our results show that the collaboration network is growing exponentially for the
last three decades, with a distribution of number of collaborators per researcher that approaches a
power-law as the network gets older. Moreover, both the distributions of number of collaborators and
production per researcher obey power-law behaviors, regardless of the geographical location or
field, suggesting that the same universal mechanism might be responsible for network growth and
productivity.We also show that the collaboration network under investigation displays a typical
assortative mixing behavior, where teeming researchers ({\it i.e.}, with high degree) tend to
collaborate with others alike. Finally, our analysis reveals that the distinctive collaboration
profile of researchers awarded with governmental scholarships suggests a strong bonus impact on
their productivity.
\end{abstract}

\maketitle

\section{Introduction}
Nowadays, scientific collaboration is understood as extremely valuable, as it integrates
skills, knowledge, apparatus and resources, allows division of labor and
the study of more difficult problems, including interdisciplinary ones. It also brings
recognition and visibility and increases the network of contacts of the researchers
involved \cite{fox1984independence,katz1997what,laudel2002what}. Scientific collaboration
is strongly correlated with production measured by publication output and other indexes
in Scientometrics \cite{beaver1978studies,lawani1986some,lee2005impact}, which has
substantially contributed to raise the interest of the scientific community in studying itself
over the last decades \cite{de1966collaboration, beaver1978studies, frame1979international,
heffner1981funded, kraut1988patterns, katz1997what}. More recently, due to the fast
growth and enormous development of the complex network science \cite{watts1998collective,
albert2002statistical, newman2002assortative, newman2003structure, ravasz2003hierarchical,
barrat2004architecture, barrat2004modeling, moreira2006competitive, lind2007spreading,
moreira2009how, galvao2010modularity, schneider2011mitigation} the subject
of scientific collaboration has been extensively studied under the framework of rather
powerful and universal paradigms \cite{barabasi2002evolution,  newman2002scientific,
newman2002scientificb, goh2002classification, newman2004coauthorship,
ramasco2004self, li2005weighted}.

The Internet and the fact that traveling became substantially less costly have
facilitated international collaborations. Still, geographical constraints
affect the dynamics of research \cite{katz1994geographical,ponds2007geographical,
pan2012world}. Different countries have different funding policies and this
impacts the publication outcome, which is correlated to collaboration.
For a country to be above the world average number of citations, it must spend more
than one hundred thousand US dollars per researcher per year \cite{pan2012world}.
At the same time, scientists with more investment in their research projects
collaborate more \cite{bozeman2004scientists}.

The social nature of collaboration might be the cause for the big disparity
in production and number of collaborators \cite{muchnik2013origins}. Inequalities
in income (Pareto
distribution \cite{pareto1897cours}) and movie co-appearance \cite{gallos2013imdb}
are examples of social distributions, characterized by a power-law profile.
For scientific collaborations, such distributions also appear, as demonstrated by
Lotka \cite{lotka1926frequency}, from the analysis of two empirical sets of publications
data in natural sciences.

Although in Lotka's analysis \cite{lotka1926frequency} only the senior authorship
has been considered, the obtained power-law was shown to be consistent with
empirical  bibliometric data taking all authors into account \cite{nicholls1986empirical}.
The so called Lotka's Law therefore seems to be valid even in different fields
than those originally considered \cite{nicholls1986empirical,pao1986empirical}.
It is also worth noting that highly prolific authors were excluded in Lotka's
procedure due to the limited number of persons in the samples. These teeming
researchers might lie outside the pure power-law distribution. Considering that
engaging in collaboration is a time consuming activity, the number of collaborators
can not be arbitrarily large, i.e., must be somehow limited. An exponential cutoff
has then been suggested as a correction to fit the distribution of productivity
\cite{newman2004coauthorship}. Measuring the distributions of citations
by city or country, a power-law distribution also arises \cite{pan2012world}, which
indicates the presence of self-similarity in the science system \cite{katz1999self}.

Nonetheless, the definition of research collaboration is problematic due to
the subjective understanding of its essential ingredients \cite{katz1997what,
laudel2002what}. This can be avoided by considering as
scientific collaboration, a research which resulted in a coauthored
scientific paper. This approach, although traditional, is not free of criticism
as there are fruitful and relevant collaborations which do not necessarily involve
a publication.
Notwithstanding, there is evidence that division of labor of theoretical or
experimental work is usually rewarded with a coauthorship \cite{laudel2002what}.
Also, analysing coauthorship makes it feasible to study collaboration of a greater
number of researchers as compared by interviewing each individual.

Despite the numerous studies about scientific production, citations and
collaborations found in the literature, it is difficult to compare
these variables as the databases used in these studies are usually unrelated.
Another problem is the small number of samples, due to a low number of respondents
in questionnaires or data used only from a specific journal. To analyse the big
picture is paramount
to work with a dense information database. Here, we used data from Lattes Platform
(http://lattes.cnpq.br), an online database maintained by CNPq (National Council
of Technological and Scientific Development), a government agency that finances
scientific research in Brazil. It contains the curricula of almost all researchers
in Brazil and their collaborators abroad, as well as information concerning their
research groups. The Lattes Curriculum
became the standard national scientific curriculum in Brazil, and compulsory for
those requiring financial support from the Brazilian government.
The curricula present detailed information concerning the researcher, including,
but not limited to, full name, gender, professional address, academic titles,
field of expertise and list of papers. Researchers are classified in 8 major fields:
Agricultural Sciences (Agr), Applied Social Sciences (Soc), Biological Sciences
(Bio), Exact and Earth Sciences (Exa), Humanities (Hum), Health Sciences (Hea),
Engineering (Eng), Linguistics and Arts (Lin), and Others (Oth). Most information
in the curriculum are provided by the researcher themselves, for example, their
list of publications.

By using this database, we may overcome some of the limitations found by other authors
\cite{newman2002scientific,barabasi2002evolution}. Due to the lack of individual
information of the researcher, the problem of author name disambiguation
\cite{newman2002scientific, tang2009bibliometric} becomes relevant, when,
for example, two or more authors share initials and surnames. This is not the
case with the Lattes Platform, where coauthorship is unambiguous and detailed
information about publications and researchers are made available by themselves.
As a consequence, this type of data allows us to study scientific
production and collaborations of individual researchers and correlations between
fields of expertise.

\section{Methodology}

The collaboration networks were built based on data of approximately 2.7 million
curricula downloaded in June 2012 from the Lattes Platform website. Files were
parsed to extract the name of the researcher, professional address and
authored papers published in
periodicals (including title, year and number of coauthors in the paper).

\begin{figure}[floatfix]
\includegraphics*[width=\columnwidth]{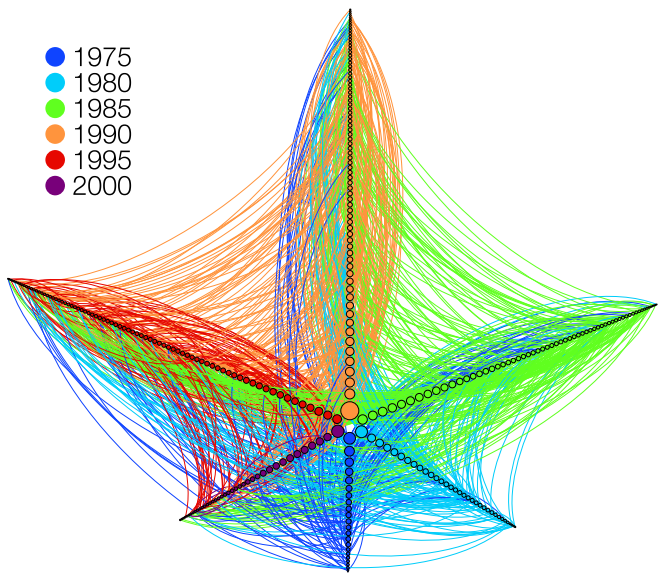}
\caption{\label{fig:sample_net}Sample network extracted from the collected data. We show links
between researchers (nodes) who  were granted a scholarship and working in fields of Medicine in the
state of S\~ao Paulo. Node size is proportional to the degree of the researcher
in the whole database.
Researchers were grouped according to the year of thier first
published paper. The first cohort (dark blue) comprises all researchers who
published their first paper before 1975. Each subsequent one, in counterclockwise
direction, comprises researchers who published within 5 years from the previous one,
up to 2000. The edges are directed, colored according to the most senior.}
\end{figure}

Due to possible typographical errors \cite{o1993characteristics}, an approximate
string matching was used to compare paper titles. We used Damereau-Levenshtein distance
\cite{wagner1975extension} as the metric and compared papers of the same year and with
the same number of authors starting with the same letter. Papers differing by
10\% or less of the maximum distance were considered to be the same paper. From
these results, we built a bipartite
network in which researchers are connected to papers. Then it is possible to eliminate
nodes (researchers or papers) from this network to study the behavior of the remaining structure.
A collaboration network is constructed by assigning an edge with a specific weight
to researchers having connections to the same paper in the bipartite network.
We show, in Fig. \ref{fig:sample_net}, a network construted from our data, using several
filters, to illustrate variety of information available.

\section{Results and discussion}
\begin{figure}[floatfix]
\includegraphics*[width=\columnwidth]{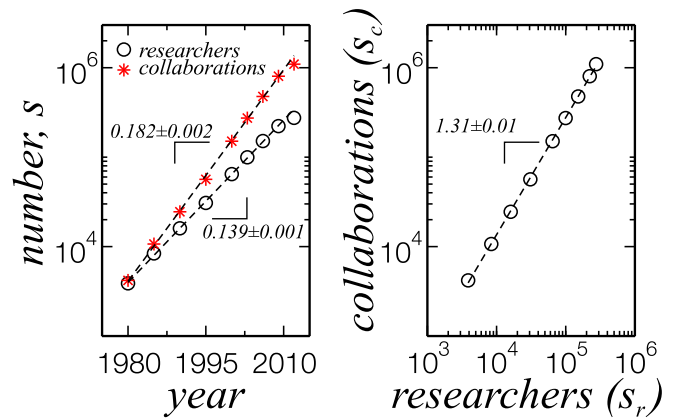}
\caption{\label{fig:growth}Left: Number of researchers with published papers (black circles)
and collaborations
between them (red stars) present in the cumulative collaboration network. Dashed lines
are exponential fits in the form $s=ae^{\alpha t}$ up to 2009, seen as straight lines
in the linear-log plot. The coefficient $\alpha$ is shown in the picture for each curve.
Deviations of the 2012 data points from the exponential fit are due to the early
acquisition of the curricula, in June of 2012. Right: Superlinear scaling of the number
of collaborations with the number of researchers. Dashed line is a power-law fit in the
form $s_c \propto s_{r}^{\lambda}$. The coefficient $\lambda$ is given by
$\lambda = \alpha_c / \alpha_r = 1.31$.}
\end{figure}

The total collaboration network (TCN) includes 275,061 researchers, with 90.4\% belonging to
the largest component. The total number of identified papers written in collaboration
is 623,984, the number of collaborations is 1,095,871 and the network comprises all
8 major areas used by the Brazilian agency CNPq to classify
researchers. The fact that more than 90\% of the network is connected is an interesting sign,
which indicates that discoveries from a field can spread in the
communities through interdisciplinary collaborations.
The extracted papers have publication date extending for several decades. By analysing
the growth of the network, we show in Fig. \ref{fig:growth} (left) that the number of
researchers ($s_r$) as well as collaborations ($s_c$) grew exponentially in the last three decades,
$s_r \propto e^{0.139t}$ and $s_c \propto e^{0.181t}$, with $t$ in years.
We also show that the number of collaborations increases superlinearly with
the number of researchers in the network. This accelerated growth has been observed
in collaboration networks \cite{barabasi2002evolution,zhang2012growing} and other
types of empirical networks \cite{dorogovtsev2002accelerated}. More recently,
it was shown that the number of social contacts and total communication also scales
superlinearly with city population size \cite{schlapfer2013scaling}.

A commendable initiative of the Brazilian government is to award scholarships to
distinguished researchers among their peers. As a matter of fact, these scholarships 
correspond to a bonus payment in addition to their base salary.
For comparison with the TCN, in Table \ref{tab:resumo} we show the
basic statistical properties of a collaboration network built with only these researchers
(scholarship network, SCN). The clustering coefficient \cite{watts1998collective},
$C$, measures the probability
that two collaborators of a given researcher have papers in common (forming
a triangle in the graph). Social networks are known to have high degree of
clustering \cite{newman2003structure}, which can be explained in terms of a hierarchical
structure \cite{ravasz2003hierarchical}. Here both networks display a high clustering
coefficient but the average value for SCN is about half of TCN. This difference
reflects the higher position in the research groups of the researchers with
scholarship. They are more likely to have contacts in other research groups,
which means being less clustered.

\begin{figure}[floatfix]
\includegraphics*[width=\columnwidth]{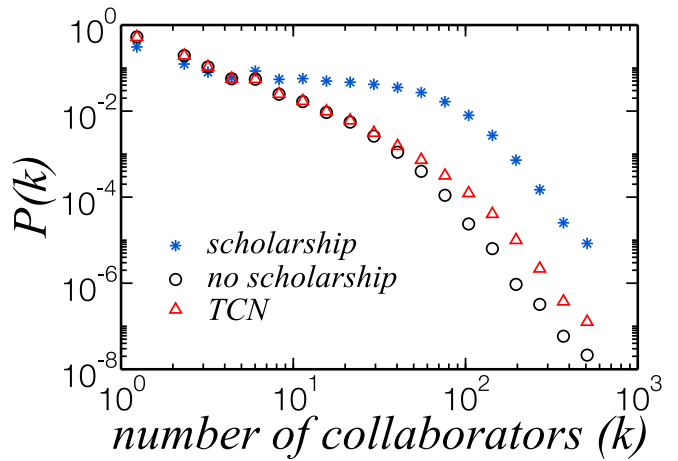}
\caption{\label{fig:pk-grant}Normalized distribution of the number of
collaborators ($k$) of researchers with scholarship (blue stars), without
(black circles) and for the TCN (red triangles). The distribution for
researchers with scholarship decreases
slowly up to one hundred collaborators, although most of them still have a small
number of collaborators. The higher proportion of researchers with high $k$ might
reflect the CNPq policy of considering the proponent's participation in research groups,
international immersion and human resources development to grant the scholarship.}
\end{figure}

A relevant question which naturally arises is how the scientific productivity and
collaboration statistics of researchers awarded with scholarships differ from regular
researchers. Studying our database, we find that researchers
in the SCN represent less than 5\% of the researchers in the TCN but contribute
with 20\% of the production. They are in average more than five times more productive,
as measured by publication output. Also, SCN is more cohesive than TCN, as measured by
the size of the giant component. To determine whether these characteristics are
cause or consequence of their scholarship is not our aim, but previous research
on collaborations strategies indicate that those with higher grants are more
likely to have more collaborators \cite{bozeman2004scientists}. The degree
distributions shown in Fig. \ref{fig:pk-grant} clearly corroborate this difference
between groups.

\begin{center}
\begin{table}
\caption{\label{tab:resumo}Statistics for the networks studied in this work.}
\begin{tabular}{l c c}
  & TCN & SCN\\
\hline
Number of researchers ($s_r$) & 275,061 & 12,302\\
Number of edges ($s_c$) & 1,095,871 & 134,186\\
Total number of papers & 623,984 & 129,699\\
Average researchers per paper & 4.51 & 5.26\\
Average papers per author ($\langle n \rangle$) & 11.1 & 61.4\\
Average number of collaborators ($\langle k\rangle$) & 8.0 & 38.1 \\
Largest component fraction & 90.4\% & 94.6\% \\
Clustering coefficient ($C$) & 0.465 & 0.266\\
Assortativity coefficient ($r$) & 0.094 & 0.230\\
\hline
\end{tabular}
\end{table}
\end{center}

\begin{figure}[floatfix]
\includegraphics*[width=\columnwidth]{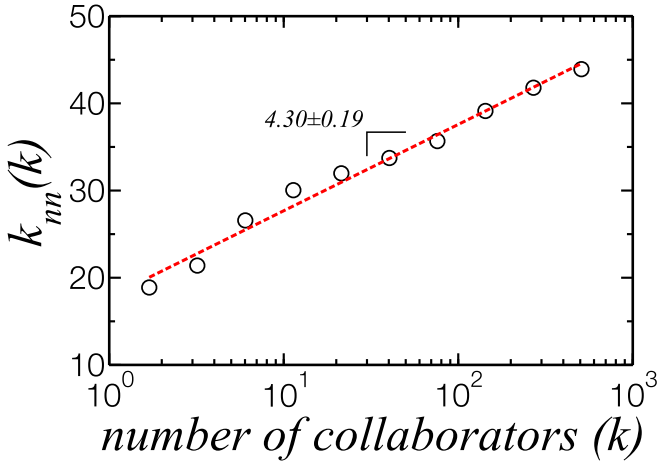}
\caption{\label{fig:knn}Variation of the average nearest-neighbor degree 
($k_{nn}$) with $k$. Being an increasing function of $k$, the network displays
assortative mixing. Researchers with high $k$ are more likely to collaborate
with other well connected researchers. This tendency, however, increases
logarithmically with $k$, as indicated by the dashed line.}
\end{figure}

The assortativity coefficient \cite{newman2002assortative},
$r$, measures the correlation between degrees of nodes at either ends of an edge.
Networks with $r<0$ are said to display disassortative mixing, while $r>0$ means
assortative mixing. Social networks, including collaborations networks, are known
to display assortative mixing \cite{newman2002assortative,barrat2004architecture}.
Another way of looking at the assortative properties of a network is through the
average nearest-neighbor degree, $k_{nn}(k)$ \cite{barrat2004modeling},
where $k$ is the number of collaborators of a researcher. This measures how well
connected the collaborators of a researcher are. If $k_{nn}(k)$ is an increasing
function, then researchers with high $k$ collaborate with other well-connected
researchers, and the network displays assortative mixing. We show
in Fig. \ref{fig:knn} that this occurs in TCN, and  that $k_{nn}$ increases
logarithmically with $k$. Assuming that researchers with a high number of
collaborators are positioned in the top of the academic hierarchy, we can infer
from Fig. \ref{fig:knn}
that prominent researchers and group leaders collaborate more among themselves.
Nonetheless, $k_{nn}$ does not grow fast but logarithmically, as
researchers growing in importance absorb the influx of new actors in the network.

\begin{figure}[floatfix]
\includegraphics*[width=\columnwidth]{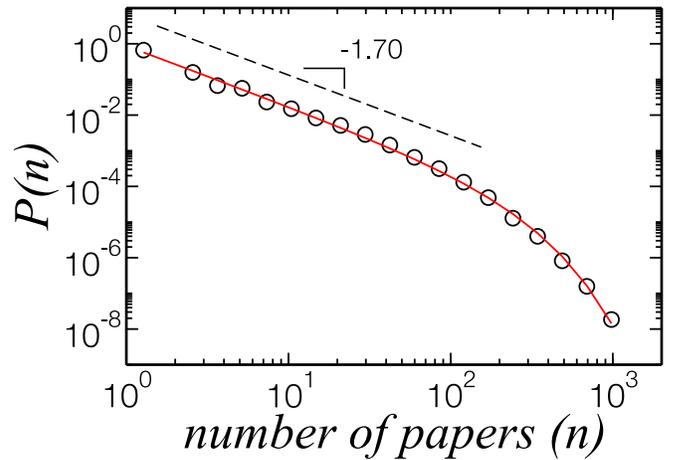}
\caption{\label{fig:prod}Distribution of scientific production of researchers
belonging to the TCN group. The solid red line is the best fit to the data points
of a power-law with exponential cutoff, $P(n)=A_{p}n^{-\beta_p}e^{-n/l_p}$,
where $\beta_p=1.70$ and $l_p=157$. The dashed black line is a power-law with
exponent $-1.70$.}
\end{figure}

It is inviting to verify if the production of researchers on Lattes Platform
obeys Lotka's Law. As shown in Fig. \ref{fig:prod}, the distribution of scientific
production (in number of papers, $n$) obeys a power-law with exponential cutoff,
$P(n)=A_{p}n^{-\beta_p}e^{-n/l_p}$, with exponent $\beta_p \approx 1.7$ and
characteristic cutoff length $l_{p}\approx 157$.

\begin{figure*}[floatfix]
\includegraphics[width=2\columnwidth]{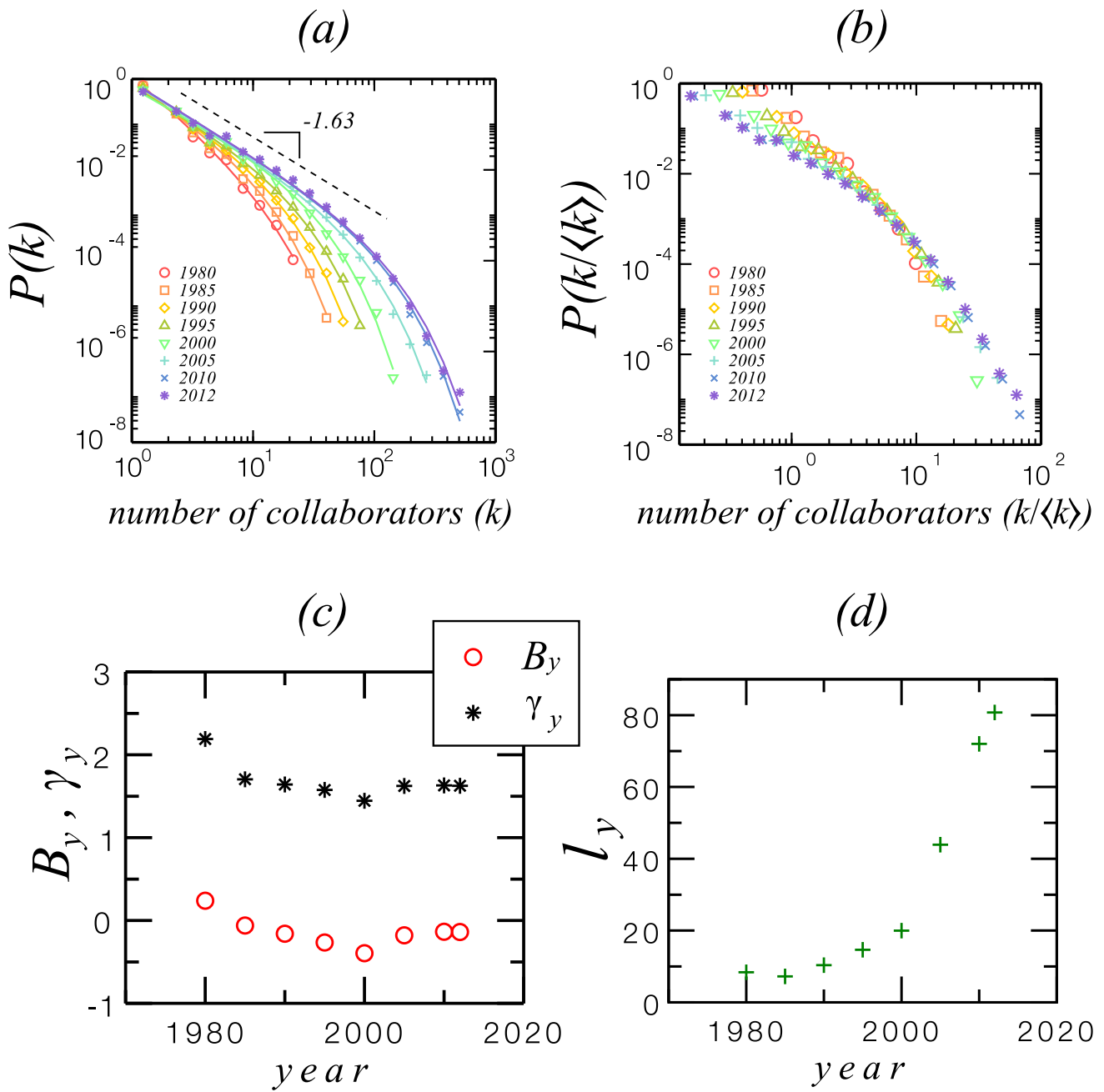}
\caption{\label{fig:pk-years}(a) Time evolution of the distribution of the number
of collaborators in the TCN. A power-law with exponential cutoff, $P(k)=B_{y}k^{-\gamma_y}e^{-k/l_y}$,
is used to fit the data points (symbols) for several years. The dashed black line
corresponds to a power-law with exponent $-1.63$. (b) Rescaling the distribution in (a)
by the relative number of collaborators for each year shows a collapse onto a single curve.
(c) The prefactors $B_y$ and
the parameter $\gamma_y$ both saturate as the network grows older. (d) The
characteristic length $l_y$ increases faster than linearly, indicating that the
distribution is approaching a power-law for an increasing number of decades.}
\end{figure*}

With this database, we can study the time evolution of the collaboration network
by analysing different groups of papers that have been published within a specific
range of years. As shown in Fig. \ref{fig:pk-years} (a), our results for the TCN indicate
that the degree distributions for different years can be all described in terms
of the same distribution used for productivity, namely, a power-law with an exponential
cutoff, $P(k)=B_{y}k^{-\gamma_y}e^{-k/l_y}$.
We show in Fig. \ref{fig:pk-years} (b)  a rescaling of these curves by the relative
number of collaborators for each year, collapsing onto a single curve.
As depicted in Fig. \ref{fig:pk-years}
(c), the values of the parameter $\gamma_y$ saturates as the network grows
older, while $l_y$ increases
faster than linearly (Fig. \ref{fig:pk-years}, d), indicating that the
distribution tends to follow a power-law for an increasing number of decades.

\begin{figure}[floatfix]
\includegraphics*[width=\columnwidth]{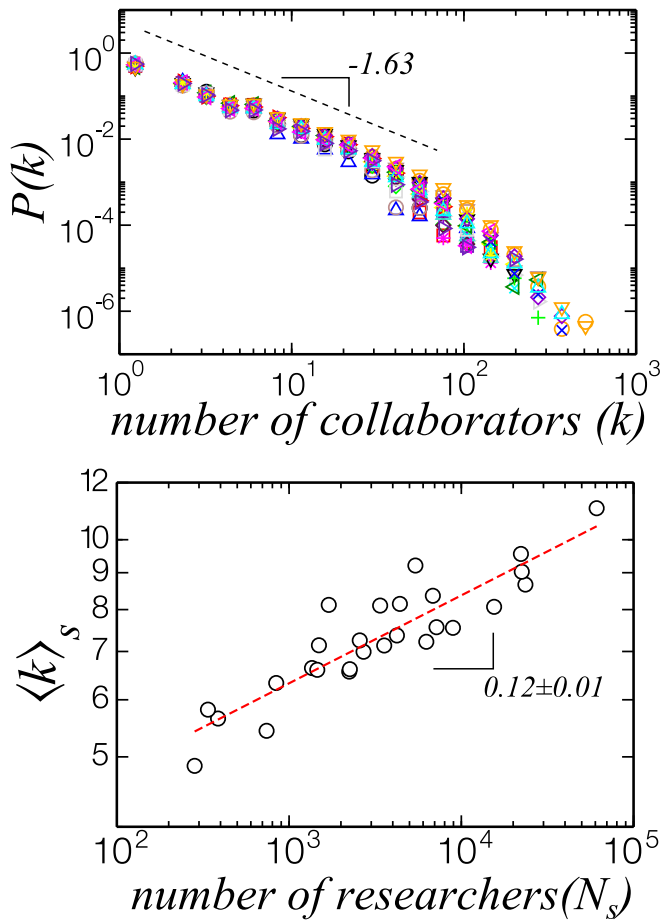}
\caption{\label{fig:pk-uf}Top: Distribution of number of collaborators
in the TCN for the 26 Brazilian states and the Federal District. The distributions
display the same behavior as the TCN (Fig. \ref{fig:pk-years}). The dashed line is a
power-law with exponent $-1.63$. Bottom: the average number of collaborators
versus the number of researchers in each state. The circles correspond to the results
for 26 Brazilian states and the Federal District. The dashed line is the best fit
of the data to a power-law $\langle k\rangle_{s} \sim N_{s}^{\delta}$,
with exponent $\delta=0.12 \pm 0.01$.}
\end{figure}

We can use the professional address information included in the curricula to
study the differences of collaboration profile due to geographical location.
As shown in Fig. \ref{fig:pk-uf} (top), the overlap of the degree distributions for the
TCN at each of the 26 states of Brazil and Bras\'ilia, the Federal District, suggests
universality in the collaboration mechanism.
The geographical location of the researcher, while not changing the shape of the
distribution, is correlated with the spectrum of the number of collaborators.
Recent allometric studies show that a large number of urban indicators
(e.g., R\&D employment, total wages, GDP, gasoline sales, length of electrical
cables) scale as a power-law of population of the city \cite{bettencourt2007growth}.
In Fig. \ref{fig:pk-uf} (bottom) we show that the average number of collaborators per
researcher in the Brazilian states $\langle k \rangle_{s}$ generally increases with their number of
researchers as a power-law, $\langle k\rangle_{s} \sim N_{s}^{\delta}$ with an exponent
$\delta=0.12 \pm 0.01$. Spatial constraints limit the number of collaborators, but,
as states are open territories, inter-state collaborations diffuses the same
mechanism throughout the whole network. Hence, individual states can be used as
representative samples of the TCN.

\begin{figure}[floatfix]
\includegraphics*[width=\columnwidth]{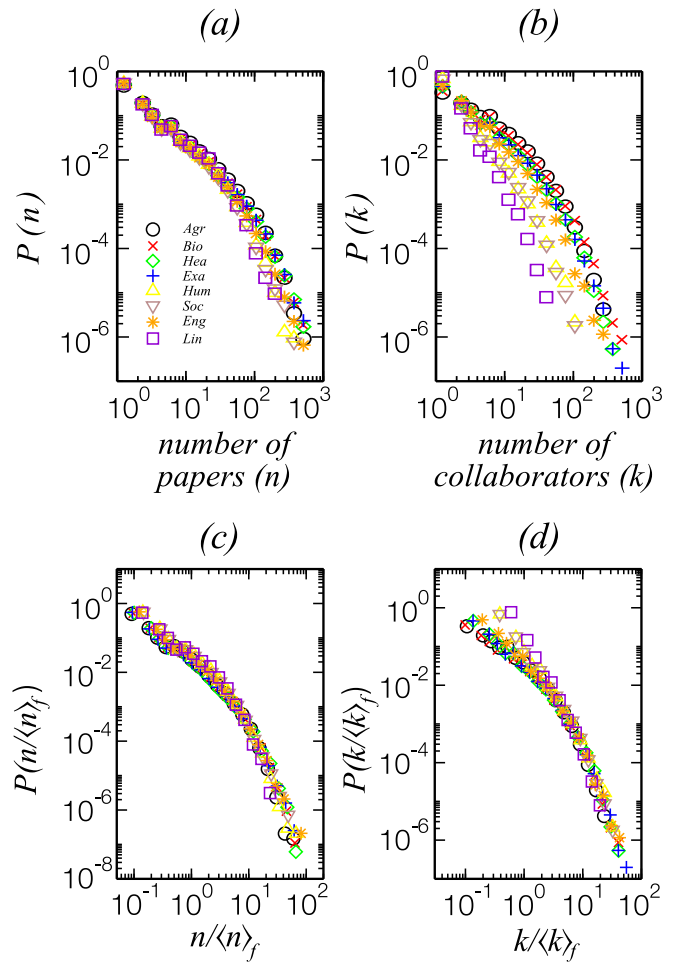}
\caption{\label{fig:scale}Distributions of the number papers published per
researcher $n$ (a), number of collaborators $k$ (b), and the respective rescaled
distributions, (c) and (d), for each of the 8 major fields. Scientists working on social
sciences and related fields (Lin, Soc and Hum) are less likely to have published more
than one hundred papers than others. They also are less likely to have more than
one hundred collaborators. Considering the average publication count 
$\langle n \rangle_f$ and average number of collaborations $\langle k \rangle_f$
in each field, all the curves collapse to a single universal behavior.}
\end{figure}

Finally, the way researchers from different fields collaborate can also be
investigated with the data downloaded from the Lattes platform. Figure
\ref{fig:scale}(a) and (b) shows that the distributions of researcher
productivity $P(n)$ as well as their corresponding degree distributions $P(k)$,
respectively, can be rather different for distinct fields. However, since different
fields are know to have different levels of productivity \cite{allison1980inequality},
by rescaling $k$ and $n$ to the corresponding average values of the field
(see Table \ref{tab:resumo2}), $\langle k \rangle_f$ and $\langle n \rangle_f$,
both $P(n)$ and $P(k)$ distributions collapse to single universal curves, as depicted
in Fig. \ref{fig:scale}(c) and (d), respectively.

\section{Conclusions}
In summary, we have used the Lattes Platform, which contains detailed and unambiguous
data of approximately 2.7 million curricula of researchers, as a database for
analysing research collaboration in Brazil. It has the advantage of displaying
individual curricula, allowing us to study collaborations in a mix of a paper-based
approach and questionnaire data.

\begin{table*}[!ht]
\caption{{\bf Statistics for researchers working on the 8 major fields associated with the TCN.}}
\begin{tabular}{l | p{2.5cm} | p{2.5cm} | p{4cm} | p{3cm}}
 & Number of researchers ($N_f$) & Researchers with scholarship ($S_f$) & Average number of papers per researcher ($\langle n \rangle_f$) & Average number of collaborators ($\langle k \rangle_f$) \\
  \hline
Agr & 31812 & 1692 & 13.9  & 11.7 \\
Bio & 39767 & 2605 & 13.1  & 12.5 \\
Hea & 67561 & 1511 & 12.6  & 9.08 \\
Exa & 33310 & 3273 & 13.5  & 9.16 \\
Hum & 26263 & 1324 & 8.90  & 3.21 \\
Soc & 20806 & 742  & 8.66  & 3.23 \\
Eng & 18365 & 1841 & 10.2  & 6.37 \\
Lin & 5202  & 300  & 9.09  & 2.06 \\
\hline
\end{tabular}
\label{tab:resumo2}
\end{table*}


We therefore built collaboration networks including all researchers data from
Lattes Platform as June 2012, and 
found that the network has grown exponentially for the last three decades.
The calculated values of the assortativity coefficient and the average
nearest-neighbor degree indicate that the networks display
assortative mixing,  where researchers having high $k$ collaborate with others
alike. Our results show that these teeming researchers are more likely to have a
scholarship and to produce more papers than researchers with low $k$. The
distribution $P(k)$ is also approaching a power-law as the network gets older.

Finally, we confirmed the validity of Lotka's Law for researchers working on different states
of Brazil and found substantial correlations between $\langle k \rangle_f$ and $N_f$. Lotka's
Law is shown to be valid for different fields: indeed, $P(n)$ and $P(k)$ follow
an universal behavior.

\section{Acknowledgments}
We thank the Brazilian Agencies CNPq, CAPES, and FUNCAP, the FUNCAP/CNPq
Pronex grant, and the National Institute of Science and Technology for
Complex Systems in Brazil for financial support.

\section{Author Contributions}

Conceived and designed the experiments: EBA AAM VF THCP JSA. Performed
the experiments: EBA. Analysed the data: EBA AAM JSA. Wrote the paper:
EBA AAM JSA.

\end{document}